\documentclass[11pt,preprint]{aastex}
\usepackage{epsfig}

\tightenlines

\newcommand	\Angstrom	{\,{\rm \AA}}

\newcommand	\beq	{\begin{equation}}
\newcommand	\cm	{\,{\rm cm}}

\newcommand	\dyn	{\,{\rm dyn}}
\newcommand	\eeq	{\end{equation}}
\newcommand	\erg	{\,{\rm ergs}}
\newcommand	\etal	{{\it et al.}}
\newcommand	\eV	{\,{\rm eV}}

\newcommand	\g		{\,{\rm g}}
\newcommand	\gtsim	{\gtrsim}		 
\newcommand	\rmH	{{\rm H}}
\newcommand	\He	{{\rm He}}
\newcommand	\HH	{{\rm H}_2}

\newcommand	\keV	{\,{\rm keV}}
\newcommand	\K		{\,{\rm K}}
\newcommand	\kms	{\,{\rm km~s}^{-1}}

\newcommand	\ltsim	{\lesssim}		 
\newcommand	\Mpc	{\,{\rm Mpc}}

\newcommand	\pc	{\,{\rm pc}}

\newcommand	\s		{\,{\rm s}}

\newlength{\figwidth}
\newlength{\figwidthb}

\countdef\decade=200
\advance\decade by \year
\countdef\hours=201
\advance\hours by \time
\divide\hours by 60
\countdef\mins=202
\advance\mins by \hours
\multiply\mins by 60
\multiply\hours by 100
\countdef\miltime=203
\advance\miltime by \hours
\advance\miltime by \time
\advance\miltime by -\mins

\begin{document}

\title{
        \vspace*{-2.0em}
        {\normalsize\rm submitted to {\it The Astrophysical Journal}}\\
        \vspace*{1.0em}
	Gamma Ray Burst in a Molecular Cloud:
	Destruction of Dust and H$_2$, and Emergent Spectrum\\
	}

\author{B.T. Draine and Lei Hao}
\affil{Princeton University Observatory, Peyton Hall, Princeton,
	NJ 08544, USA;\\ 
	{\sf draine@astro.princeton.edu, haol@astro.princeton.edu}}

\begin{abstract}
A gamma ray burst with strong optical-UV emission
occuring in a molecular cloud will photodissociate
H$_2$, photoionize H$_2$, H, and He, and destroy dust grains.
We model these processes, including time-dependent radiative transfer in
both continuum radiation and the resonance lines of $\HH$.
The UV will pump $\HH$ into 
vibrationally-excited levels.
We calculate the absorption spectrum imprinted on radiation from
the GRB at various times.
In addition to the strong absorption lines of $v=0$ $\HH$ 
at $\lambda < 1110\Angstrom$
due to cold ambient gas,
we find that radiation reaching us from the GRB and its afterglow
will show strong absorption lines due to vibrationally-excited
$\HH$ at $1110 < \lambda < 1705\Angstrom$.
These absorption lines, if observed, would provide unequivocal evidence
for association of the GRB with molecular gas.
Low-resolution spectra will exhibit conspicuous features due to
clustering of individual lines; a list of the strongest such absorption 
features is given for spectral resolution $R\approx350$ characteristic
of the grism on the Swift UV-Optical Telescope.
\end{abstract}


\section{Introduction\label{sec:intro}}

Gamma-ray bursts (GRBs) rank among the most dramatic, most energetic, and least
understood celestial phenomena.  
Although at least some GRBs are now known
to originate from cosmological distances, the progenitor objects, or their
locations in galaxies, are not yet known.
Only $\sim$40\% of GRBs have detectable optical afterglows, whereas the X-ray
afterglow detection rate is $\sim$100\%;  
the undetected optical afterglows may have been extinguished by dust in the
host galaxy (see Lazzati, Covino, \& Ghisselini 2001; Ramirez-Ruiz, Trentham,
\& Blain 2001; and references therein).
Since GRBs may be associated with star-forming regions 
(Paczy\'nski 1998, 1999; MacFadyen, Woosley, \& Heger 2001), 
it is of interest to consider 
phenomena which would be indicative of
molecular gas in the vicinity of the GRB.

At least some GRBs are accompanied by
intense optical emission, as demonstrated by detection of
a 9th magnitude optical transient coinciding with
GRB\,990123 (Akerlof \etal\  1999) and optical afterglows associated
with other GRBs (e.g., GRB\,990510: Stanek \etal\ 1999; Israel \etal\ 1999).
The energy radiated in the optical-UV flash and afterglow 
can be substantial, and
can have dramatic effects on interstellar
gas and dust in the vicinity of the GRB.
The $h\nu > 13.6\eV$ emission will photoionize the
gas, and
the pulse of optical radiation will vaporize
dust grains out to substantial distances from the GRB
(Waxman \& Draine 2000, hereafter WD00).
The gamma-rays and hard X-rays emitted by the GRB will 
contribute to ionization of the nearby gas, but the UV and
soft X-rays have the dominant effect 
because of the much greater number of photons,
and much larger photoabsorption and photoionization cross sections.
The time-dependence of atomic and ionic absorption lines and photoionization
edges in the spectra of GRBs has been discussed by
Perna \& Loeb (1998),
B\"ottcher et al (1999),
and
Lazzati, Perna, \& Loeb (2001).
In the present work we focus on the rich absorption spectrum of $\HH$.

Ultraviolet radiation will destroy $\HH$
in the vicinity of the GRB, but before destruction some of the
$\HH$ molecules will be vibrationally excited 
by UV pumping.
A preliminary study of the UV pumping (Draine 2000) found that high
degrees of vibrational excitation would prevail in the $\HH$
undergoing photodissociation and photoionization, and an 
estimate of the column density of this vibrationally-excited
gas indicated that it could
imprint a conspicuous absorption spectrum on light reaching us
from the GRB flash and afterglow.
If observed, this absorption at $\lambda \ltsim 1650\Angstrom$
would be a clear sign that the GRB was in close proximity
to a molecular cloud.

The previous study did not explicitly include self-shielding in the
$\HH$ lines, or the details of the competition between photoionization
and photodissociation of the $\HH$, and therefore it was only possible
to make approximate estimates of the column density of vibrationally-excited
$\HH$.
In the present work we carry out a detailed calculation of the
effects of a high-luminosity optical transient on surrounding dust and
molecular gas.  
We calculate the evolution of the dust and gas in the
neighborhood of the source of radiation, including radiative transfer
through the absorbing medium.
We include the dominant sources
of opacity: dust grains, $\HH$, $\HH^+$, H, He, and $\He^+$.
Computational limitations dictate using radial zones which can be
optically thick to 13.6~eV -- 100~eV ionizing photons, and in the Lyman and
Werner band transitions of $\HH$.
We describe a numerical method which can be used to follow the motion
of the ionization and dissociation fronts in this regime.

The GRB light curve adopted for the calculation is given in \S\ref{sec:flash},
where it is compared with the observed light curve of GRB 990123.
Because GRB 990123 appears to have been unusually bright, in this paper
we consider the effects of an optical transient with a peak ultraviolet 
luminosity less than 1/40 that of GRB 990123, and an integrated optical
luminosity per steradian perhaps 1\% of GRB 990123.
The computational method is described in \S\ref{sec:radiative_transfer},
and the physical processes (dust destruction, photoexcitation, and
photoionization) are summarized in \S\ref{sec:physical_processes}.

In \S\ref{sec:results} we present results for our 
adopted lightcurve $L_\nu (t)$, for two 
cloud densities ($n_\rmH=10^3$ and $10^4\cm^{-3}$) and
for two different power-law indices ($\beta=-0.5$ and -1) for
the UV-X-ray spectrum $L_\nu \propto \nu^\beta$.
In \S\ref{sec:obs_prospects} we discuss the prospects for observing these
absorption features in the spectrum of a GRB fireball or afterglow, either
with ground-based observations of GRBs at redshift $z>1.25$, or using
the grism on the Swift GRB Explorer for GRBs at redshift $z > 0.06$.
We summarize our results in \S\ref{sec:summary}.

\begin{figure*}[ht]	
\centerline{\epsfig{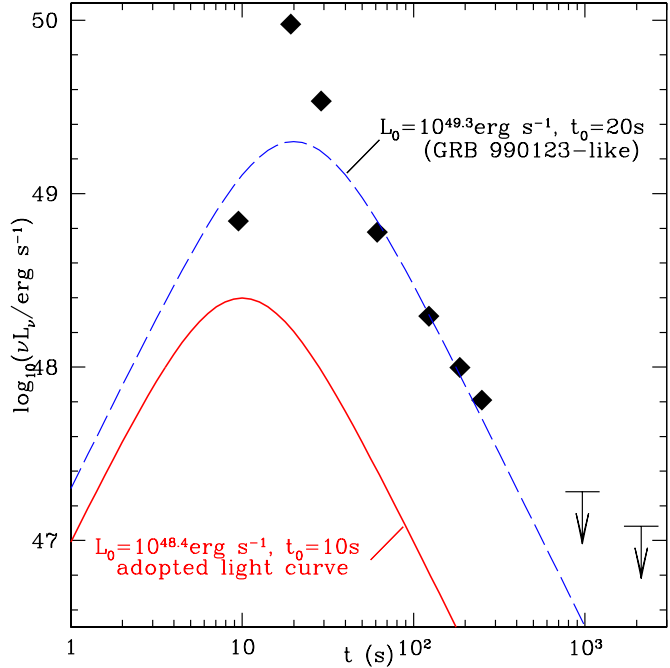}}
\figcaption{\label{fig:L(t)}
	{\footnotesize
	Filled symbols: observed light curve of GRB 990123 
	(Akerlof et al. 1999) at
	$\lambda\approx 2000\Angstrom$ in the rest frame
	(assuming 
	$H_0=65\kms\Mpc^{-1}$, an open cosmology
	with $\Omega=0.3$, and zero extinction).
	Broken curve: eq.\ (\ref{eq:L(t)}) with
	$L_0=2\times10^{49}\erg\s^{-1}$ and $t_0=20\s$.
	This underestimates the energy in the optical transient of
	GRB 990123 optical transient.
	Solid curve: eq.\ (\ref{eq:L(t)}) with
	$L_0=2.5\times10^{48}\erg\s^{-1}$ and $t_0=10\s$, which
	we adopt for the optical transient of a typical GRB.}
	}
\end{figure*}

\section{The Optical-UV Flash
	\label{sec:flash}
	}

Let $L \equiv 4\pi dP/d\Omega$
where 
$dP/d\Omega$ is the radiated power per solid angle in our direction.
It is not yet known what should be taken for 
the ``typical'' $L(t)$ for a GRB at optical -- soft X-ray wavelengths.

In Figure \ref{fig:L(t)} we show $\nu L_\nu(t)$ of 
GRB 990123, at redshift $z=1.6$,\footnote{%
	Assuming 
	$H_0=65\kms\Mpc^{-1}$, an open cosmology
	with $\Omega=0.3$, and zero extinction.
	}
at rest-frame wavelength $\sim 2000\Angstrom$.

As a simple fitting function to describe the time-dependent luminosity,
we adopt
\beq
\nu L_\nu = L_0 \frac{4(t/t_0)^2}{[1+(t/t_0)^2]^2} 
\left(\frac{h\nu}{13.6\eV}\right)^{1+\beta} ~~~.
\label{eq:L(t)}
\eeq
A spectral index $\beta=-0.5$ is suggested by simple
models for the reverse shock (Waxman \& Draine 2000).
Steeper spectral indices have been observed for optical-UV afterglows:
for GRB 010222, at redshift $z\geq 1.475$ (Masetti et al.\ 2001) 
Lee et al.\ (2001) find $\beta=-0.90$ at $(1+z)t=1.7\times10^4\s$, 
Stanek et al.\ (2001) find $\beta=-1.07$ at $(1+z)t=1.7\times10^4\s$,
and
Masetti et al.\ (2001) find $\beta=-0.55$ for 
$8\times10^4\ltsim(1+z)t\ltsim 1.6\times10^5\s$.
 
With $L_0=2\times10^{49}\erg\s^{-1}$, $t_0=20\s$ and $\beta\approx-0.5$,
eq.\ (\ref{eq:L(t)}) approximates the light curve of
GRB 990123 at $50\ltsim t \ltsim 500\s$, 
although falling at least a factor 5 below
the observed peak bightness at $t=20\s$ (see Fig.\ \ref{fig:L(t)}).

The $L_\nu\propto t^{-2}$ dependence of equation (\ref{eq:L(t)}) at
$t \gg t_0$ is required to approximate the observed rapid decline in brightness
of GRB 990123 between 20 and 300 sec (see Fig.\ \ref{fig:L(t)}).
However, at later times afterglows appear to fade less rapidly: 
for GRB 010222, 
$L_\nu \propto t^{-0.75\pm0.05}$ for 
$1.7\times10^4\ltsim(1+z)t\ltsim3\times10^4\s$,
and $L_\nu \propto t^{-1.3}$ for $8\times10^4\ltsim (1+z)t\ltsim 3\times10^6\s$
(Stanek et al.\ 2001; Masetti et al.\ 2001).
For the present study we are concerned with the lightcurve at
$t\ltsim 10^3\s$, 
which we will approximate using equation (\ref{eq:L(t)}).

GRB 990123 was in the top 0.4\% in gamma-ray fluence for
GRBs detected by the BATSE instrument, and
LOTIS observations further suggest that most GRBs detected by BATSE 
have lower ratios of
optical to gamma-ray emission (Williams et al.\ 1999).
Accordingly, we adopt $L_0= 2.5\times10^{48}\erg\s^{-1}$ -- 
i.e., a peak luminosity about $1/40$ of what was observed for
GRB 990123 -- and
$t_0\approx10\s$ as ``typical'' parameters.
For $\beta=-0.5$ this corresponds to the optical transient having
an energy between 1 and 13.6 eV given by
$E_{\rm OT} = 2\pi L_0 t_0 [1 - (1/13.6)^(1+\beta)]/(1+\beta)
= 1.57\times10^{50}\erg$;
for $\beta=-1$, 
$E_{\rm OT} = 2\pi L_0 t_0 \ln 13.6
= 4.1\times10^{50}\erg$.

\section{Radiative Transfer
	\label{sec:radiative_transfer}
	}

We divide the cloud into spherical shells $j=1,...,N$ of uniform
thickness $\Delta R$,
with outer radii $R_{j}$
and midpoints 
$\bar{R}_j\equiv R_{j}-0.5\Delta R$.
At each radius $r$, let $t_r\equiv t-r/c$ be the ``retarded'' time.

The shell thickness $\Delta R$ is small enough 
($n_\rmH\Delta R < 10^{21}\cm^{-2}$)
so that an individual
shell is optically thin in dust, but the shells may be optically thick
to photoionizing radiation, or in absorption lines of $\HH$.  
We therefore calculate rates for
photoionization, photodissociation, or photoabsorption
averaged over the shell volume 
\beq
\Delta V_j\equiv (4\pi/3)\left[R_j^3-(R_j-\Delta R)^3\right] ~~~.
\eeq

At frequency $\nu$, the optical depth contributed by shell $j$ is
\beq
\label{eq:Deltatau}
\Delta\tau_j(\nu)\equiv
\delta\tau_{d,j}(\nu) + 
\delta\tau_{\HH,j}(\nu) + 
\delta\tau_{\HH^+,j}(\nu) + 
\delta\tau_{\rmH,j}(\nu) + 
\delta\tau_{\He^0,j}(\nu) + 
\delta\tau_{\He^+,j}(\nu) +
{\sum_\alpha}^\prime \delta\tau_{\alpha,j}~~
\eeq
where
$\delta\tau_{d,j}$, 
$\delta\tau_{\HH,j}$, 
$\delta\tau_{\HH^+,j}$,
$\delta\tau_{\rmH,j}$, 
$\delta\tau_{\He^0,j}$, and 
$\delta\tau_{\He^+,j}$ are the contributions in shell $j$
due to extinction by dust, and continuous absorption by 
$\HH$, $\HH^+$, $\rmH$,
$\He^0$, and $\He^+$, respectively;
$\tau_{\alpha,j}$ is the contribution of an individual $\HH$ absorption
line $\alpha$ (see \S\ref{sec:line_absorption}).
Absorption by metals not in dust grains 
(initially O, N, Ne, and S will be most important, and C, Mg, Si, and Fe 
will enter the gas phase
in regions where
grains are vaporized)
has been
neglected in eq.\ (\ref{eq:Deltatau}), because the present calculation is
limited to the gas phase species 
H, H$^+$, H$_2$, H$_2^+$, He, He$^+$, and He$^{2+}$.
We have thereby underestimated the attenuation at 
$h\nu\gtsim 100$eV.

We divide the electromagnetic spectrum into 3 intervals:
$h\nu < 11.1\eV$, $11.1 < h\nu < 13.6\eV$, and $13.6 < h\nu < 20 \keV$.
For purposes of modelling the heating of dust grains by the UV-optical
radiation,
we include the contribution of $\HH$ absorption lines only within the
interval $11.1 < h\nu < 13.6\eV$; the summation ${\sum}^\prime$
in equation (\ref{eq:Deltatau})
is restricted to lines $\alpha$ within this interval, with $\delta\tau_\alpha$
calculated from eq.\ (\ref{eq:delta_tau_alpha}) below.

The total optical depth from $r=0$ to $R_j$ is
\beq
\label{eq:tau}
\tau_j(\nu,t_r)=\sum_{i=1}^{j}\Delta\tau_i(\nu,t_r) ~~~.
\eeq

\subsection{Absorption and Scattering by Dust}

\begin{figure*}[ht]	
	\centerline{\epsfig{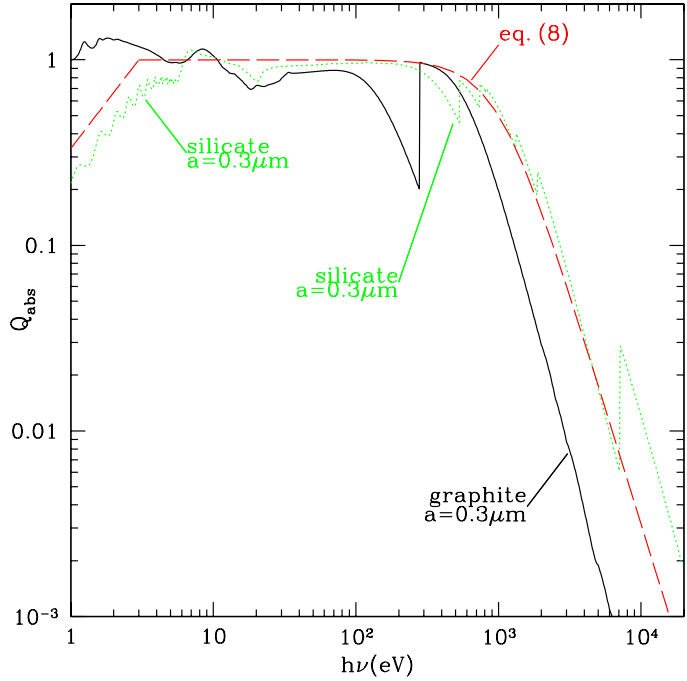}}
\figcaption{\label{fig:qabs}
	\footnotesize
	Absorption efficiency $Q_{\rm abs}$
	for $a=0.3\micron$ radius graphite and
	silicate grains.
	}
\end{figure*}

The contribution by dust in shell $j$ is
\beq
\delta\tau_{d,j}(\nu) = 
n_{d,j} \pi a_j^2 (Q_{\rm abs}(\nu)+Q_{\rm sca}(\nu))\Delta R ~~~,
\eeq
where $a_j(t_r)$ is the dust radius in shell $j$ at retarded time $t_r$,
and $Q_{\rm abs}(\nu)$ and $Q_{\rm sca}(\nu)$ 
are the usual efficiency factors for absorption and scattering.
The radiation from the GRB is concentrated in a shell-like
``pulse'' with a spatial thickness of $\sim 10$ light-sec 
(see Fig.\ \ref{fig:L(t)}).
Scattering is important because it reduces the intensity in the pulse;
the scattered photons will fall behind the outward-propagating
pulse unless the scattering angle is very small.
For scattering angle $\theta\ll 1$, after travelling a distance $\delta R$
the scattered photon will fall
behind the unscattered pulse by a time
\beq
\delta t \approx \frac{\delta R\,\theta^2}{2c} = 
1.4\s \frac{\delta R}{10^{18}\cm}\left(\frac{\theta}{\rm arcmin}\right)^2 ~~,
\eeq
so that the time delay will be significant for scattering angle
$\theta\gtsim 5 {\rm arcmin}$.
The dust grains will have radii $a\approx3\times10^{-5}\cm$ 
(see \S\ref{sec:initial_conditions}).
For $a=0.25\micron$, the characteristic scattering angle
is $\sim3$~arcmin at 2 keV, and $\sim10$~arcmin at 0.5 keV
(see Figs. 3 and 4 of Smith \& Dwek 1998)
so
we take the scattering efficiency factor to be
\beq
Q_{\rm sca}(\nu)=\left\{ \begin{array}
{l@{\quad\quad}l}
0 & {h\nu}<1\eV ~,\\ 
1 & 1\eV\le h\nu<1\keV ~,\\ 
0 & h\nu\ge 1\keV ~.
\end{array} \right.
\eeq
In Fig.\ \ref{fig:qabs} we show $Q_{\rm abs}$ calculated for
$a=0.3\micron$ graphite and silicate grains.
At $h\nu\gtsim 1\keV$, the silicate absorption efficiency is much larger
because of the inner shell absorptions from Mg, Si, and Fe.
In the present calculation we do not distinguish between grain types,
and we need a simple estimate for the absorption efficiency which
can be rapidly evaluated.
As a simple approximation, we adopt
\beq
\label{eq:qabs}
Q_{\rm abs}(\nu)=\left\{ \begin{array}
{l@{\quad\quad}l}
(h\nu/3\eV) & {h\nu}<3\eV ~,\\ 
1/\left[ 1 + (h\nu/\keV)^{2.5}\right] & h\nu\ge 3\eV ~,
\end{array} 
\right.
\label{eq:Qabs}
\eeq
which can be seen in Fig. \ref{fig:qabs} to approximate the
average absorption cross sections of graphite and silicate grains.

When grain destruction has reduced the grain radius to 
$a\ll 3\times10^{-5}\cm$,
eq.\ (\ref{eq:Qabs}) will overestimate $Q_{\rm abs}$; when this is true,
however, grain destruction is nearly complete and the error in
overall grain destruction will be unimportant.

\subsection{Continuous Absorption by Atoms, Molecules, and Ions}

Continuous absorption by species $X$ contributes
\beq
\delta\tau_{X,j}(\nu) = n(X_j)\Delta R \sigma_X(\nu) ~,
\eeq
where $n(X_j,t_r)$ is the abundance of $X$ 
in shell $j$ at retarded time $t_r$,
and $\sigma_X(\nu)$ is the continuous absorption cross section.

The probability that a photon which enters the shell will be absorbed 
or scattered is
$(1-e^{-\Delta\tau_j})$.  
For a homogeneous shell, a fraction
$\Delta\tau_{X,j}/\Delta\tau_j=n_{X,j}\sigma_X\Delta R/\Delta\tau_j$ 
of the absorptions will be due to 
continuous absorption by species $X$ (e.g., $X+h\nu\rightarrow X^++e^-$).
The photon absorption rate per $X$,
averaged over the shell, is
\beq
\zeta_{X}=\frac{\Delta R}{\Delta V_j}
\int_{\nu_X}^\infty 
\frac{L_\nu d\nu}{h\nu} 
e^{-\tau_{j-1}}
\frac{(1-e^{-\Delta \tau_j})}{\Delta\tau_j}
\sigma_{X}(\nu)~~~.
\label{eq:zeta^pi}
\eeq
Integration over frequency $\nu$ is accomplished numerically, with
$\Delta\tau_j$ and $\tau_{j-1}$ evaluated using equations (\ref{eq:Deltatau})
and (\ref{eq:tau}).

\subsection{Line Absorption
	\label{sec:line_absorption}}

Line absorption is important for $\HH$.
Consider a specific absorption line $\alpha$, with central frequency
$\nu_\alpha$.
We assume that the $\HH$ has a Maxwellian velocity distribution
with broadening parameter $b$, and we compute the dimensionless
equivalent width
\beq
W_\alpha(R_j) = \int \frac{d\nu}{\nu}
\left\{1-\exp\left[-N_\alpha(R_j)\sigma_\alpha(\nu)\right]\right\} ~,
\eeq
where $\sigma_\alpha(\nu)$ is the (Voigt profile) absorption cross
section for absorption line $\alpha$, and
\beq
N_\alpha(R_j) = \sum_{i=1}^{j}
\left[n\left(\HH(v_\alpha,J_\alpha)\right)\right]_i\Delta R
\eeq
is the column density of the vibration-rotation state $(v_\alpha,J_\alpha)$
producing absorption line $\alpha$.

The incremental equivalent width contributed by shell $j$ is
\beq
\delta W_{\alpha,j} = W_\alpha(R_j)-W_\alpha(R_{j-1}) ~.
\eeq
If we approximate the line absorption as due to many lines
randomly distributed over the
frequency interval $[\nu_{\min},\nu_{\max}]$, the optical
depth contribution due to line $\alpha$ averaged over the wavelength
interval is (see Draine \& Bertoldi 1996)
\beq
\delta\tau_{\alpha,j} = 
\frac{\delta W_{\alpha,j}}{\ln[\nu_{\max}/\nu_{\min}]} ~~~{\rm for~}\nu_\alpha
\in [\nu_{\min},\nu_{\max}]
\label{eq:delta_tau_alpha}
\eeq
and the mean photoabsorption rate 
in radial shell $j$ for excitation out of level $X_\alpha$ by
the absorption line $\alpha$ is
\beq
\zeta_{X\alpha,j} = 
e^{-\tau_{j-1}}\frac{L_\nu ~\delta W_{\alpha,j}}{4\pi \bar{R}_j^2 N_\alpha(R_j) h} ~~.
\label{eq:zeta_alpha}
\eeq
Strong line absorption by $\HH$ occurs primarily in 
$[h\nu_{\min},h\nu_{\max}]=[11.2,13.6]\eV$,
so we include the optical depth correction (\ref{eq:delta_tau_alpha})
only for this interval.

\section{Physical Processes
	\label{sec:physical_processes}
	}

\subsection{Dust Destruction by Thermal Sublimation}

The thermal 
sublimation rate from a grain at temperature $T$ 
can be approximated by (Guhathakurta \& Draine 1989,
hereafter GD89)
\beq
{da\over{dt}}=
-\left ({\frac{m} {\rho}}\right )^{1\over 3} \nu_0 e^{-B/kT} ~~~,
\eeq
where $m$ is the mean atomic mass, and
$B$ is the chemical binding energy per atom. 
We take
$\nu_0=1\times10^{15}{\rm s}^{-1}$, $B/k=7\times10^4{\rm K}$,
and $\rho/m = 1\times 10^{23}{\rm{cm}}^{-3}$ as 
representative values for refractory grains 
(GD89;
WD00).
We assume that all of the energy of absorbed photons (including X-rays)
is converted to heat -- we neglect the energy of photoelectrons and X-ray
fluorescence, but this is a smal fraction of the total absorbed energy.

In the middle of shell $j$, 
the grain temperature $T$ is determined by
(WD00)
\beq
\int_0^{\infty}
\frac{L_\nu d\nu}{4\pi \bar{R}_j^2}
\exp\left(-\tau_{\nu,j-1}-0.5\Delta\tau_{\nu,j}\right)
Q_{abs} \pi a^2 =
\langle Q \rangle_T 4\pi a^2 \sigma T^4 -4\pi a^2 \frac{da}{dt} \frac{\rho}{m} B \quad,
\eeq
where
$\langle Q\rangle_T$ is the Plank-averaged absorption efficiency. 
For the temperature range of interest for dust sublimation, 
$2000\K\lesssim T\lesssim 3000\K$,
we approximate (WD00)
\beq
\langle Q\rangle_T \approx
\frac{0.1 (a/10^{-5}\cm) (T/2300{\rm K})}{1+0.1(a/10^{-5}\cm)(T/2300{\rm K})}
~~~,
\label{eq:Q_T}
\eeq
intermediate between the emissivities of astronomical silicate and
graphite.
%

\subsection{Dust Destruction by Electrostatic Stresses?}

Waxman \& Draine (2000) noted that
a highly-charged dust grain 
can also be destroyed by ``Coulomb explosion'' if the electrostatic
stress exceeds the tensile strength of the grain materials, when the
grain potential reaches
\beq
U_{\rm frac}\approx 3000 {\rm V}
(S_{\max}/10^{11}\dyn\cm^{-2})^{1/2}(a/10^{-5}\cm) .
\eeq
The importance of this process is controversial.
Waxman \& Draine argued that submicron grains would likely have high enough
tensile strengths ($S_{\max} > 1\times10^{11} \dyn\cm^{-2}$)
that highly-charged 
grains would not undergo fission by this process, but would
instead be gradually eroded by the
process of ``ion field emission'', where singly-charged ions are emitted 
one-by-one
when the grain potential exceeds a critical value 
\beq
U_{\rm IFE}\approx 3\times10^3 {\rm V}\left(a/10^{-5}\cm\right)
\eeq
(Muller \& Tsong 1969; Draine \& Salpeter 1979).
Fruchter, Krolik, \& Rhoads (2001), on the other hand, have argued
for lower tensile strengths ($S_{\max} \ltsim 10^9 \dyn\cm^{-2}$), 
in which case grain fission would
take place when the grain potential reached 
$U_{\rm frac}\approx 
300 {\rm V} (a/10^{-5}\cm) (S_{\max}/10^9\dyn\cm^{-2})^{1/2}$.
Fruchter et al argue that ``the large flux of energetic photons
bombarding the grain is likely to damage the
grain's crystalline structure'', reducing the tensile strength.

While chemical bonds will undoubtedly be disrupted by ionization,
it seems likely that chemical bonds will be promptly reestablished
in the warm grain -- i.e., ``annealing'' will take place.
In this case, a grain which has already been charged
to the ion-field emission threshold potential
$U_{\rm IFE}$ would emit one ion per escaping photoelectron.
To sustain ion field emission, the grain must emit photoelectrons
with sufficient energy to overcome the potential
$U_{\rm IFE} = 3000 V (a/10^{-5}\cm)$.
Consider, for example, a grain with radius $a=3\times10^{-5}\cm$, charged to
$U_{\rm IFE}\approx 9~{\rm kV}$:
only photoelectrons with 
kinetic energy at the grain surface $E > eU \approx 9 \keV$
are able to escape.  
For a grain with composition MgFeSiO$_4$, a photon of energy
$h\nu = 10 \keV$ (for example) has a total absorption coefficient
$\alpha = 600\cm^{-1}$ in the grain material, but only $\sim4\%$ of
the absorptions produce photoelectrons with $E > 9 \keV$,\footnote{
	$h\nu=10\keV$ photons absorbed by
	the Mg, Si, and Fe K shells (with photoelectric thresholds $E_T=$ 
	1.32, 1.87, and 7.11 keV) produce photoelectrons with
	$h\nu-E_T =$ 8.68, 8.13 and 2.89 keV, respectively.}
for an effective absorption coefficient $24\cm^{-1}$.
The total electron density in the grain is $1\times10^{24}\cm^{-3}$,
so the effective photoabsorption cross section per electron is
$\sim 24\times10^{-24}\cm^{-3}$ -- a factor of 24 larger than estimated
by Waxman \& Draine (2000), but a factor $\sim10^3$ smaller than the
absorption cross section just above the Si K edge which Fruchter et al
used to estimate the rate of photoelectric emission driving grain
fragmentation by electrostatic stresses.

Grain destruction by ion field emission requires a fluence
$F_{\rm IFE} \approx (\rho/m)/\alpha_{\rm IFE}$,
where $\rho/m\approx 10^{23}\cm^{-3}$ is the atomic density and
$\alpha_{\rm IFE}$ is the effective absorption coefficient for
producing photoelectrons with energies $> eU_{\rm IFE}$.
For $U_{\rm IFE}=9$kV, we have $\alpha_{\rm IFE}=24\cm^{-1}$ and
$F_{\rm IFE}=4\times10^{21}\cm^{-2}$ of $h\nu\gtsim 10\keV$ photons.
For $\beta=-0.5$ and distance $D$, the fluence of $h\nu>10\keV$ photons is
$2.2\times10^{21}(L_0t_0/10^{49.4}\erg)(\pc/D)^2\cm^{-2}$ (neglecting
absorption), so we might expect grain destruction by ion field emission
at distances $D\ltsim 1 \pc$.  
However, as we will see below (\S\ref{sec:results}),
grain destruction by thermal sublimation is effective out to distances
of several pc.


\subsection{Photoexcitation of H$_2$}

Let index $i=1,...,N$ denote the vibration-rotation levels of 
the electronic ground state X$^1\Sigma_g^+$.
At a given point in the cloud, let $p_i(t)\equiv 2n(\HH(v_i,J_i))/n_\rmH$ be the 
fraction of H nuclei which are in vibration-rotation level $i$.

Prior to the GRB, $\HH$ is almost entirely in the first two or three 
rotational levels ($J$=0,1,2) of the $v$=0 vibrational level
of the electronic ground state (X$^1\Sigma_g^+$). 
The first UV photons to be absorbed will photoexcite the $\HH$ to the 
B$^1\Sigma_u^+$ and C$^1\Pi_u^{\pm}$ electronic states
(via Lyman or Werner band transitions), which requires $h\nu>11.2$ and 
$12.3\eV$, respectively) 
or photoionize the $\HH$ to $\HH^+$, which requires $h\nu>15.4\eV$
[for $\HH(v=0)$].
$\HH$ which is photoexcited to the B or C states will decay back to the 
ground electronic state in $\sim10^{-9}\s$, 
but typically to a vibrationally-excited level 
(e.g.,$v=5$) with the rotational quantum number changed by 
$\Delta J =0,\pm 2$. 
The lifetimes of the vibrationally-excited levels are long compared 
to the timescale for photoexcitation or photoionization, 
so depopulation of the vibrationally-excited levels of $\HH$ 
will be primarily by UV photoexcitation and photoionization.

The intensity of the radiation field at 1000\AA\ relative to the
local interstellar radiation field estimate of Habing (1968) is 
\beq
\chi = \frac{(\nu u_\nu)_{1000\Angstrom}}{4\times 10^{-14}\erg\cm^{-3}}
= 7\times10^{12}
\frac{\nu L_\nu}{10^{48}\erg\s^{-1}}
\left(	\frac{\pc}{R}	\right)^2
\exp[-\tau(1000\AA)].
\eeq

Stimulated emission in the UV transitions is negligible if $\chi\ll10^{19}$. 
We therefore assume that photoexcitation out of level $i$ of X$^1\Sigma_g^+$ to vibration-rotation states of B$^1\Sigma_u^+$ and C$^1\Pi_u^{\pm}$ will be followed (immediately) by spontaneous decay either to bound levels $j$ of the ground electronic state X$^1\Sigma_g^+$, or else to the vibrational continuum of X$^1\Sigma_g^+$ (i.e., photodissociation).
$\HH$ formation on grains or in the gas will be neglected since we will be concerned with conditions where the $\HH$ is destroyed in a matter of seconds or minutes. 
Collisional deexcitation can be neglected if $n_{\rm H}/{\rm cm}^{-3}\ll\chi$. 
The spontaneous decay via quadrupole transitions out of vibrationally-excited level $i$ can also be neglected if $\chi\gg10^4$.


Let $T_{ji}$ be the rate for photopumping of $\HH$
out of level $i$ into level $j$,
and let $\zeta_i^{(pd)}$ and $\zeta_i^{(pi)}$ be the rates for
photodissociation and photoionization out of level $i$.
If 
\beq
\label{eq:T_ii}
T_{ii}\equiv
-\left[\sum_{j\neq i} T_{ji} + \zeta_i^{(pd)} + \zeta_i^{(pi)}\right] ~,
\eeq
then the $\HH$ abundances $p_i\equiv 2n[\HH(v_i,J_i)]/n_\rmH$
evolve according to
\beq
\label{eq:dpi/dt}
\frac{dp_i}{dt} = \sum_j T_{ij}p_j ~.
\eeq
Following Draine \& Bertoldi (1996), 
we consider the 299 bound states of $\HH$ with $J\leq 29$, and
calculate equivalent widths $W_\alpha$
using Lyman and Werner band oscillator strengths
from Abgrall \etal\ (1993a,b) and Roueff (1993).

Let $\alpha$ denote a Lyman or Werner band transition from a specific 
vibration rotation state
$i$ of the electronic ground state to an excited state $u(\alpha)$.
Photoexcitation rates $\zeta_\alpha$ in shell $j$ are calculated using 
eq.\ (\ref{eq:zeta_alpha}).
Let $P_{j,u}$ be the probability that electronically-excited state 
$u$ will decay to vibration-rotation state $j$ of the electronic ground
state.
Averaged over the shell, the rate for UV pumping from vibration-rotation
level $i$ to vibration-rotation level $j$ is
\beq
T_{ji}={\sum_\alpha}^\prime P_{j,u(\alpha)}\zeta_\alpha
\eeq
where the summation is restricted to transitions $\alpha$ out of
level $i$.
The photodissociation rate out of level $i$ is
\beq
\zeta_i^{(pd)} = {\sum_\alpha}^\prime p_{diss}[u(\alpha)]
\eeq
where $p_{diss}(u)$ is the probability that level $u$ will decay
by a transition to the vibrational continuum.

We neglect line absorption by the $\HH^+$ ion, as it is likely to be
of secondary importance.

\subsection{$\HH + h\nu \rightarrow \HH^+ + e^-$}

The different vibrationally-rotation levels of the electronic ground state 
of $\HH$ will have different energy thresholds for photoionization.
We will, however, neglect these differences and assume that all 
$\HH$ in the ground electronic state requires
$h\nu > 15.4\eV$ to be photoionized,
with the cross section as given by eq.(17-19) of Yan et al. (1998).
The photoionization rate $\zeta_{\HH}^{(pi)}$ averaged over the shell
is evaluated using eq.\ 
(\ref{eq:zeta^pi}).

\subsection{$\HH^+ + h\nu \rightarrow \rmH^+ + \rmH$}

$\HH^+$ has 19 vibrational levels; 
for photoionization of $\HH(v=0)$ the $\HH^+$ vibrational distribution is
expected to peak at
$v=2$, with 80\% of the population in $v\leq5$ (Busch \& Dunn 1972).
With radiative lifetimes $\sim10^7\s$, there will be negligible
deexcitation by spontaneous decay during the optical transient.
Averaged over the $\HH^+$ vibrational distribution, the 
photodissociation cross section
found by von Busch \& Dunn (1972) can be fit by
\beq
\sigma
= 2.7\times10^{-16}\cm^2 \left(\frac{h\nu}{29\eV}\right)^2
\left(1-\frac{h\nu}{29\eV}\right)^6 ~~~
{\rm for}~h\nu<29\eV,
\label{eq:sigma(H2+pd)}
\eeq
giving a peak cross section $\sigma\approx3\times10^{-18}\cm^2$
at $h\nu\approx7.3\eV$ ($\lambda\approx1700\Angstrom$).
The photodissociation rate $\zeta_{\HH^+}^{(pd)}$ averaged over
the shell is evaluated using eq.\ (\ref{eq:zeta^pi}).

\subsection{$\HH^++h\nu \rightarrow 2\rmH^+ + e^-$}

For photoionization of $\HH^+$ we adopt a cross section
\beq
\sigma
= 9.3\times10^{-19}\cm^2 \left(\frac{h\nu}{15.4\eV}\right)^{-2}
~~~{\rm for}~h\nu>15.4\eV
\eeq
which approximates the photoionization cross section calculated by
Bates \& Opik (1968) for the ground vibrational state of $\HH^+(v=0)$.
The photoionization rate $\zeta_{\HH^+}^{(pi)}$ averaged over the
shell is evaluated using eq.\ (\ref{eq:zeta^pi}).

\subsection{$\rmH + h\nu \rightarrow \rmH^+ + e^-$}

Atomic $\rmH$ from the photodissociation 
of $\HH$ or $\HH^+$
can be photoionized by photons with 
$h\nu > I_\rmH$ with
a cross section (Osterbrock 1989)
\beq
\sigma
=
6.30\times 10^{-18}\cm^2 \left[
1.34\left(\frac{h\nu}{ 13.6\eV}\right)^{-2.99} -
0.34\left(\frac{h\nu}{13.6\eV}\right)^{-3.99}
\right] ~~~{\rm for}~h\nu>13.6\eV.
\eeq
The photoionization rate $\zeta_{\rmH}^{(pi)}$ 
averaged over the shell is evaluated using
eq.\ (\ref{eq:zeta^pi}).
\subsection{$\He + h\nu \rightarrow \He^+ + e^-$ and $\He^+ + h\nu \rightarrow
\He^{2+} + e^-$}
The cross sections for photoionization of He and $\He^+$
are (Osterbrock 1989)
\beq
\sigma
=7.83\times 10^{-18}\cm^2
\left[ 1.66\left (\frac{h\nu}{24.6\eV}\right )^{-2.05}-
0.66\left (\frac{h\nu}{24.6\eV}\right )^{-3.05}\right]
~~~{\rm for}~h\nu>24.6\eV
\eeq
\beq
\sigma
=1.58\times 10^{-18}\cm^2
\left[ 1.34\left (\frac{h\nu}{54.4\eV}\right )^{-2.99}-
0.34\left (\frac{h\nu}{54.4\eV}\right )^{-3.99}\right]
~~~{\rm for}~h\nu>54.4\eV
\eeq
The photoionization rates $\zeta_{\rm He}^{(pi)}$ and
$\zeta_{\rm He^+}^{(pi)}$
averaged over the shell are evaluated using eq.\ (\ref{eq:zeta^pi}).

\begin{figure*}[ht]	
\centerline{\epsfig{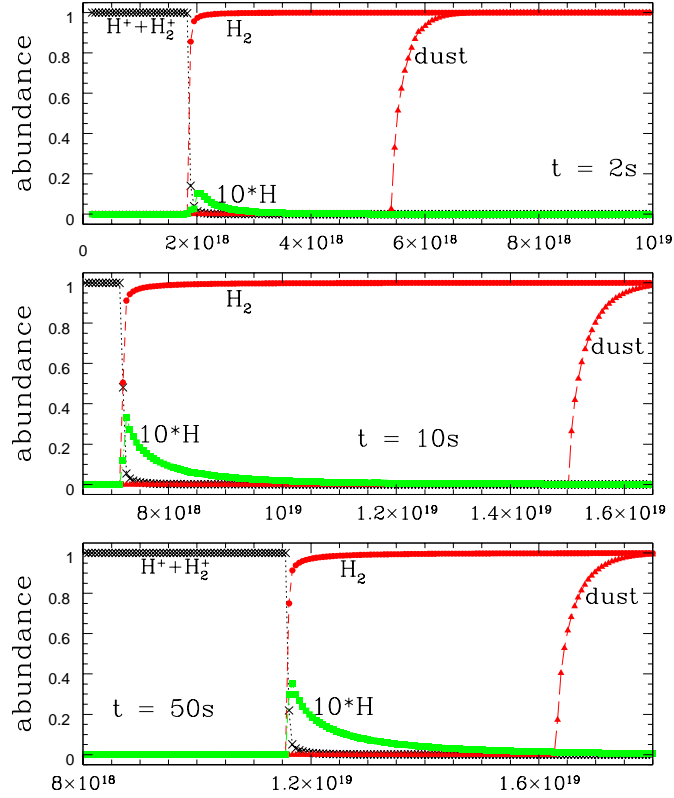}}
\vspace*{-0.8cm}
\figcaption{\label{fig:abund_1e3}
	{\footnotesize
	Snapshots of the ionization/dissociation/dust destruction
	fronts at three values of the retarded time $t_r$ for our standard GRB 
	[$L_\nu \propto \nu^{-1/2}$ given by eq.\ (\ref{eq:L(t)}) with
	$L_0=2.5\times10^{48}\erg\s^{-1}$, $t_0=10\s$, 
	$\beta=-0.5$]
	in a cloud with $n_\rmH=10^3\cm^{-3}$.
	Open triangles: dust mass $(a/a_i)^3$;
	circles: $2n(\HH)/n_\rmH$;
	open diamonds: $10\times n(\rmH)/n_\rmH$;
	filled diamonds: $n(\rmH^+)/n_\rmH$.
	}}
\end{figure*}

\begin{figure*}[ht]	
\centerline{\epsfig{
	file=f4.cps,
	width=\figwidth}}
\vspace*{-0.8cm}
\figcaption{\label{fig:abund_1e4}
	{\footnotesize
	Same as Fig.\ \ref{fig:abund_1e3}, but for $n_\rmH=10^4\cm^{-3}$.
	}
	}
\end{figure*}

\section{Photochemical Evolution
	\label{sec:evolution}
	}
\subsection{Initial Conditions \label{sec:initial_conditions}}

As initial conditions we take He to be neutral, and 100\% of the
hydrogen in $\HH$, divided equally between the first two rotation levels:
\beq
2\frac{n[\HH(v\!=\!0,J\!=\!0)]}{n_\rmH}=
2\frac{n[\HH(v\!=\!0,J\!=\!1)]}{n_\rmH}=0.5 ~~~.
\eeq

We assume that dust grains
of initial radius $a_i=3\times10^{-5}\cm$ 
initially contribute a mass equal to 1\% of
the H mass, with initial number density 
\beq
n_{d} = \frac{0.01 n_\rmH  m_\rmH} {(4\pi/3)\rho a_i^3} ~~~,
\eeq
where $\rho \approx 2 \g \cm^{-3}$ is the assumed 
density of the grain material, 
$n_\rmH$ is the number density of $\rmH$ nuclei, and
$m_\rmH$ is the mass of one $\rmH$ atom.
Thus the initial geometric cross section per H atom is
\beq
\frac{n_{d}}{n_\rmH} \pi a^2 = 2.1\times10^{-22}\cm^2\rmH^{-1} ~~~.
\eeq

Since the radiation propagates only in the outward direction
(we are neglecting scattered light),
the time evolution of each shell depends only on interior shells.
We therefore calculate the complete time evolution for each shell,
beginning with shell 1 and proceeding outward.
For each shell $j$, we begin integrating at retarded time $t_r=0$. 
For each value of $t_r$ we use stored information for
interior shells $i < j$ to compute the time- and frequency-dependent
optical depth $\tau_{j-1}$ due to material interior to shell $j$.

The calculation for shell $j$ is terminated 
when $t_r > 100 t_0 = 10^3\s$ [see eq.\ (\ref{eq:L(t)})]
and nearly all of the flash energy has been expended, {\it or}
if {\it all} of the following conditions
are fulfilled:
1. all $\HH$ has been photodissociated to $2\rmH$ and then 
photoionized into $2\rmH^+$, or
photoionized first to $\HH^+$ and then to $2\rmH^+$
[our criteria for this are $ \sum_i p_i(v_i,J_i)< 10^{-6}$, $x(\rmH)<10^{-6}$,
and $x(\HH^+)<10^{-6}$];
2. the dust has been destroyed ($a=0$);
3. helium is fully photoionized to $\He^{2+}$.

\subsection{Integration Scheme}

A very simple explicit scheme is used to advance the abundances in
each shell:
\beq
y(t+\Delta t)=y(t)+(dy/dt)_t\Delta t ~~~,
\eeq
where $y$
is $a$, $n(\rmH)$, $n(\rmH^+)$, $n(\He^0)$, $n(\He^+)$, $n(\He^{2+})$,
or $\HH$ vibration-rotation level population $p_i$.

For all cases we used radial zone thickness $\Delta R=6\times10^{16}\cm$.
At a given distance $r$ and retarded time $t_r$,
we take the time step $\Delta t$ to be
\beq
\Delta t = \min[\epsilon_d t_d, \epsilon_p t_p]
\eeq
\beq
t_d^{-1} =
\max\left[
-T_{ii},
\zeta_{\rmH}^{(pi)},
(\zeta_{\HH^+}^{(pi)}+\zeta_{\HH^+}^{(pd)}),
\zeta_{\rm He}^{(pi)},
\zeta_{{\rm He}^+}^{(pi)},
\right] ~~~,
\eeq
\beq
t_p^{-1} = \zeta_{\HH}^{(pi)}\frac{n(\HH)}{n(\HH^+)} ~~~;
\eeq
$t_d$ is the shortest time scale for destruction,
while $t_p$ is the timescale for production of $\HH^+$.
The rates $T_{ii}$ are given by equation (\ref{eq:T_ii}).
We used
$(\epsilon_d,\epsilon_p)=(0.2,0.2)$ for all cases; we verified that the
computed results were insensitive to factor-of-two variations in $\epsilon_d$ 
and $\epsilon_p$. 

\section{Results
	\label{sec:results}
	}

Radial abundance profiles are shown
in Fig.\ \ref{fig:abund_1e3},
for three values of the retarded time $t_r$, for
an optical transient with the light curve given by eq.\ (\ref{eq:L(t)})
with spectral index $\beta=-0.5$,
located in a molecular cloud of density $n_\rmH=10^3\cm^{-3}$.

In Fig.\ \ref{fig:abund_1e4} we show abundance profiles for the same
three values of retarded time, 
for the same optical transient but situated in a cloud of density 
$n_\rmH=10^4\cm^{-3}$.
In its initial molecular form, 
an individual shell is opaque to ionizing radiation
$h\nu \ltsim 70\eV$, and is highly opaque in individual
$\HH$ resononance lines; it is therefore essential to use the shell-averaged
rates given by eq.\ (\ref{eq:zeta^pi}) for continuum absorption,
or (\ref{eq:zeta_alpha}) for line absorption.
As one would expect, the increased opacity of the medium delays propagation
of the destruction fronts.

\begin{figure*}[ht]	
	\centerline{\epsfig{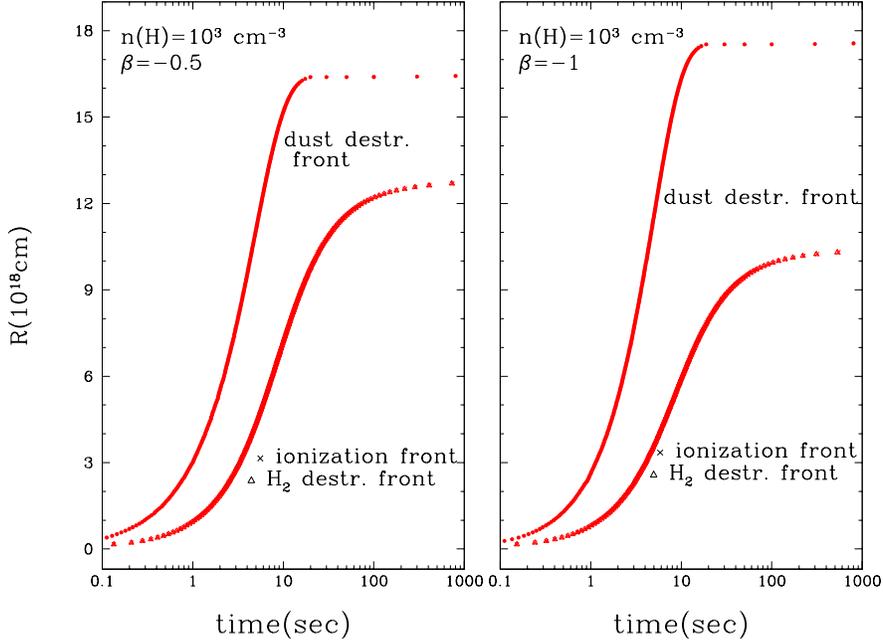}}
\figcaption{\label{fig:fronts_1e3}
	{\footnotesize
	Location of the dust destruction front, $\HH$ destruction front, and
	ionization front as functions of time for our model optical
	transient in molecular gas with $n_\rmH=10^3\cm^{-3}$, for 
	$\beta=-0.5$ and -1.
	The ionization front and $\HH$ destruction front are nearly
	coincident.
	}}
\end{figure*}

\begin{figure*}[ht]	
\centerline{\epsfig{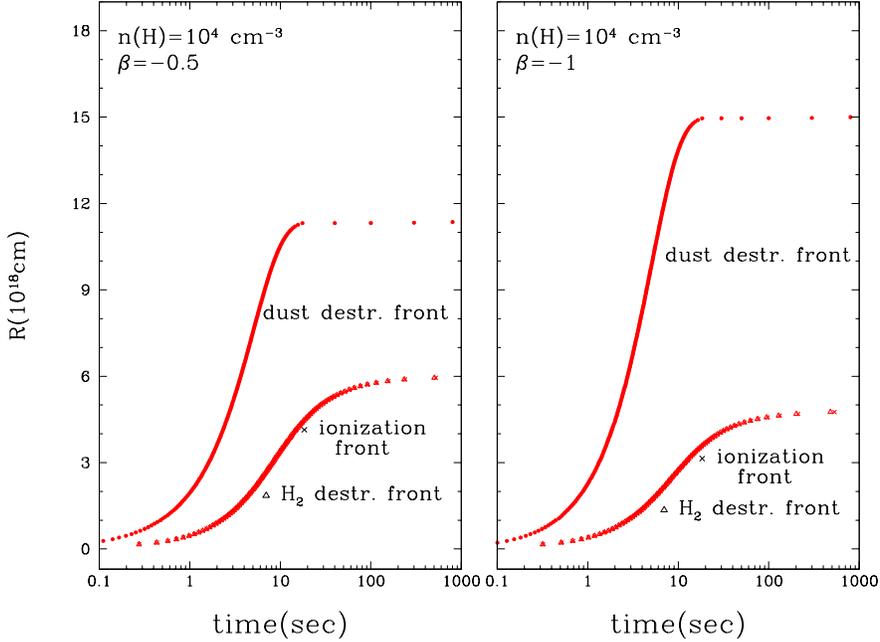}}
\figcaption{\label{fig:fronts_1e4}
	{\footnotesize
	As in Fig.\ \ref{fig:fronts_1e3}, but for cloud density
	$n_\rmH=10^4\cm^{-3}$.
	}}
\end{figure*}

In Figs.\ \ref{fig:fronts_1e3} and \ref{fig:fronts_1e4} we show the
locations of the dust destruction, dissociation, and ionization fronts
as functions of time for our standard optical transient in clouds
of density $n_\rmH=10^3$ and $10^4\cm^{-3}$, and we also show the
front locations as functions of time for a GRB with a softer
UV spectrum characterized by $\beta=-1$.

\subsection{Dust Destruction
	\label{sec:dust_destruction}
	}

In Fig.\ \ref{fig:abund_1e3} 
we see that by $t_r=2\s$ the dust destruction front has already
reached $R\approx3.7\times10^{18}\cm$, significantly ahead of the
ionization/dissociation front at $\sim1.9\times10^{18}\cm$.
At late times the dust destruction front ``stalls'' when the transient
luminosity has dropped to the point where it can no longer heat dust at this
distance to
the $T\approx3000\K$ temperatures required for significant sublimation,
while the ionization/dissociation front continues to advance at a rate
proportional to the luminosity.
As a result, for this case the ionization/dissociation front 
catches up to the dust destruction front (at $t\approx70\s$) 
and moves past it at late times
(see Fig.\ \ref{fig:fronts_1e3}).
The dust destruction front (see Fig.\ \ref{fig:abund_1e3}) has a thickness
$\delta R \approx 1\times10^{18}\cm$, corresponding to a gas column
density $\sim\!10^{21}\cm^{-2}$, and an optical-uv 
extinction (prior to dust destruction)
$\sim 0.4$.

In Figs.\ \ref{fig:fronts_1e3} and \ref{fig:fronts_1e4} 
we show the location of the dust destruction
front as a function of retarded time $t_r$ for our adopted optical
transient (for spectral indices $\beta=-1$ and $\beta=-0.5$)
and cloud densities $n_\rmH=10^3$ and $10^4\cm^{-3}$.
For both cases we confirm the conclusion of WD00: 
the dust destruction front initially advances well
ahead of the ionization/dissociation front, with the dust heated primarily
by $h\nu\ltsim11\eV$ radiation which is not absorbed by H, He, or
$\HH(v=0)$.
For $n_\rmH=10^3\cm^{-3}$ and $\beta=-0.5$ the dust destruction radius
$R_d \approx 1.2\times10^{19}\cm$ for our adopted optical transient,
while for $10^4\cm^{-3}$ we find $R_d\approx7.7\times10^{18}\cm$.

For a transient with peak luminosity $3\times10^{48}\erg\s^{-1}$
in 1--7eV photons -- close to the peak 1--7.5~eV luminosity 
$2.4\times10^{48}\erg\s^{-1}$ for our adopted light curve with $\beta=-0.5$ --
WD00 estimated the grain destruction radius to be
$R_d\approx1.4\times10^{19}\cm$ for $n_\rmH\ltsim10^3\cm^{-3}$, and
$1.3\times10^{19}\cm$ for $n_\rmH=10^4\cm^{-3}$.
The grain destruction radius estimated by WD00 for these two cases 
is in good agreement with our detailed modelling.

\subsection{Photodissociation and Photoionization
	\label{sec:h2_destruction}
	}
In Fig.\ \ref{fig:abund_1e3} we can see that the dissociation front
and ionization front are merged (i.e., they do not separate).
However, the ionization front is relatively sharp (with the fractional
ionization varying from 0.1 to 0.9 over a column density
$\Delta N_\rmH\approx10^{18}\cm^{-2}$), whereas the dissociation front
(zone where H atoms are present)
tends to be broader, and to extend ahead of the ionization front.
For $n_\rmH=10^3\cm^{-3}$ (see Fig.\ \ref{fig:abund_1e3})
the dissociation front extends $\sim1\times10^{17}\cm$,
corresponding to the column density
$n_\rmH\delta R\approx 10^{20}\cm^{-3}$ required for $\HH$ to absorb most
of the photons between 11.1 and 13.6~eV (Draine \& Bertoldi 1996).
As expected, the ionization front is unresolved, since the column
density of a single shell ($5\times10^{19}\cm^{-2}$) corresponds to an
optical depth of $\sim35$ at the Lyman edge [hence the importance of
using eq.\ (\ref{eq:zeta^pi}) to obtain the average
rate of photoionization in an optically-thick shell].
This is even more pronounced for the $n_\rmH=10^4\cm^{-3}$ case, for which
we have used the same shell thickness $\Delta R$ as for $n_\rmH=10^3\cm^{-3}$.

\begin{figure*}[ht]		
	\centerline{\epsfig{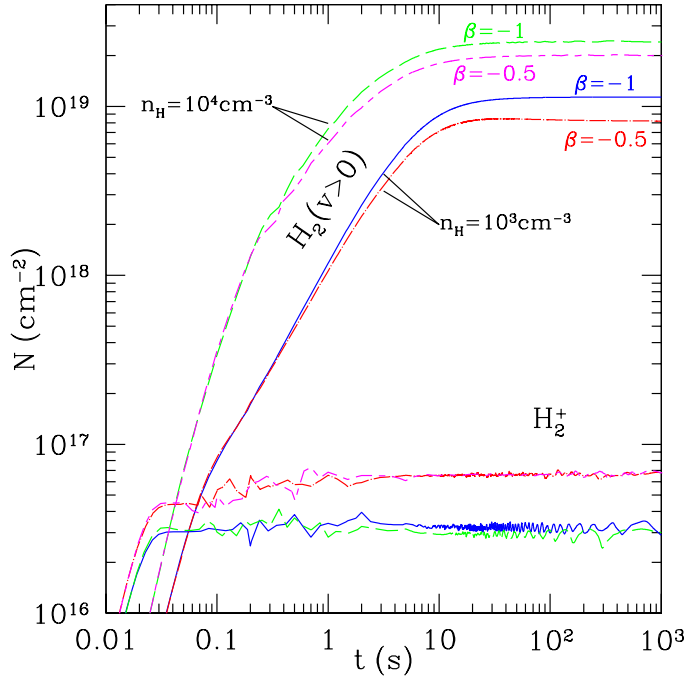}}
	\figcaption{
	\label{fig:N(H2star)} 
	{\footnotesize 
	Column densities of H$_2^+$ and vibrationally-excited H$_2$
	for $n_\rmH=10^3\cm^{-3}$ and $10^4\cm^{-3}$, and $\beta=-1$
	and -0.5.  
	The oscillations in $N(\HH^+)$ are due to
	inaccuracies in the numerical method as the ionization front
	traverses a radial zone.  
	The column density of $\HH^+$
	quickly saturates at $\sim 3-10\times10^{16}\cm^{-2}$, but the
	column density of vibrationally-excited $\HH$ continues to
	rise until the ultraviolet flash fades.
	}}
\end{figure*}
\begin{figure*}[ht]		
	\centerline{\epsfig{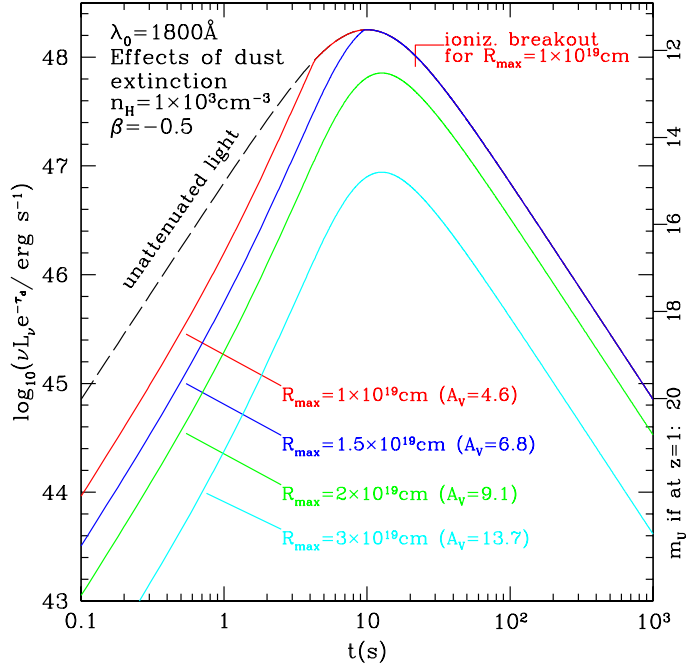}}
\figcaption{\label{fig:extincted_lightcurve}
	{\footnotesize
	Dust-attenuated light curves at rest wavelength 
	$\lambda_0=1800\Angstrom$ for different values of the
	cloud radius $R_{\max}$, for 
	a GRB with $\beta=-0.5$ in a cloud with $n_\rmH=10^3\cm^{-3}$.
	A dust extinction cross section/H of $4.2\times10^{-22}\cm^2$
	is assumed.
	For a GRB at $z=1$, the U band apparent magnitude scale is
	indicated at right.
	}
	}
\end{figure*}

\begin{figure*}[ht]	
	\centerline{\epsfig{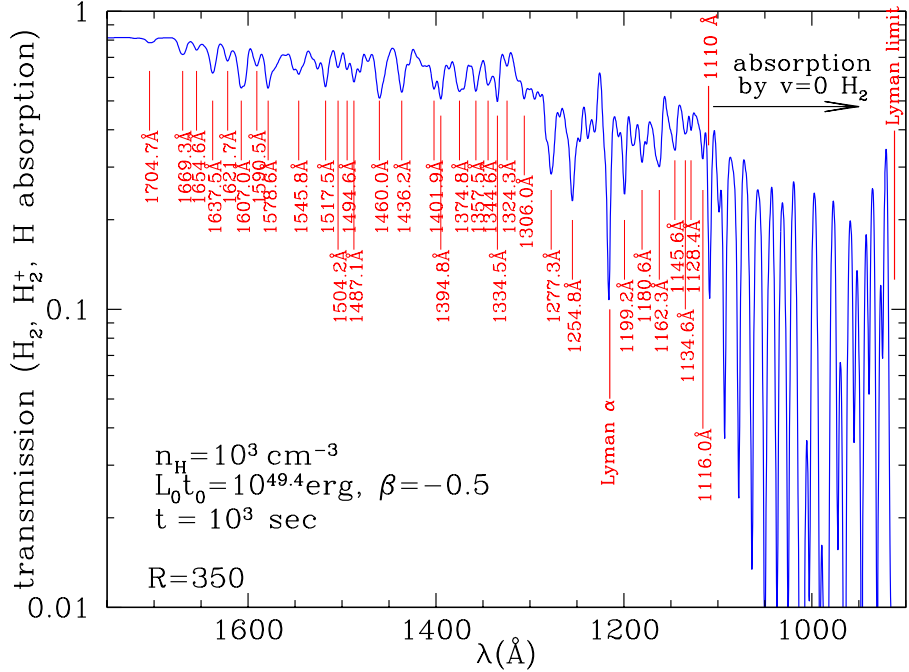}}
\figcaption{\label{fig:tran_n3_b5_t1000_R350}
	{\footnotesize
	Transmission spectrum of the intervening H$_2$, H, and H$_2^+$
	at selected times, for a GRB in a cloud with $n_\rmH=10^3\cm^{-3}$,
	as observed by a spectrograph with $R=\lambda/{\rm FWHM}_\lambda=350$.
	The GRB is assumed to have a spectrum with $\beta=-0.5$
	(i.e., $L_\nu\propto\nu^{-0.5}$).
	Continuum absorption by dust contributes additional
	attenuation.
	At resolution $R=350$ a number of $\lambda > 1110\Angstrom$ 
	absorption ``features'' from vibrationally-excited $\HH$ are
	evident, and are labelled by wavelength.
	Some of the stronger lines contributing to these features
	are listed in Table \ref{tab:features}.
	}
	}
\end{figure*}

\begin{figure*}[ht]		
	\centerline{\epsfig{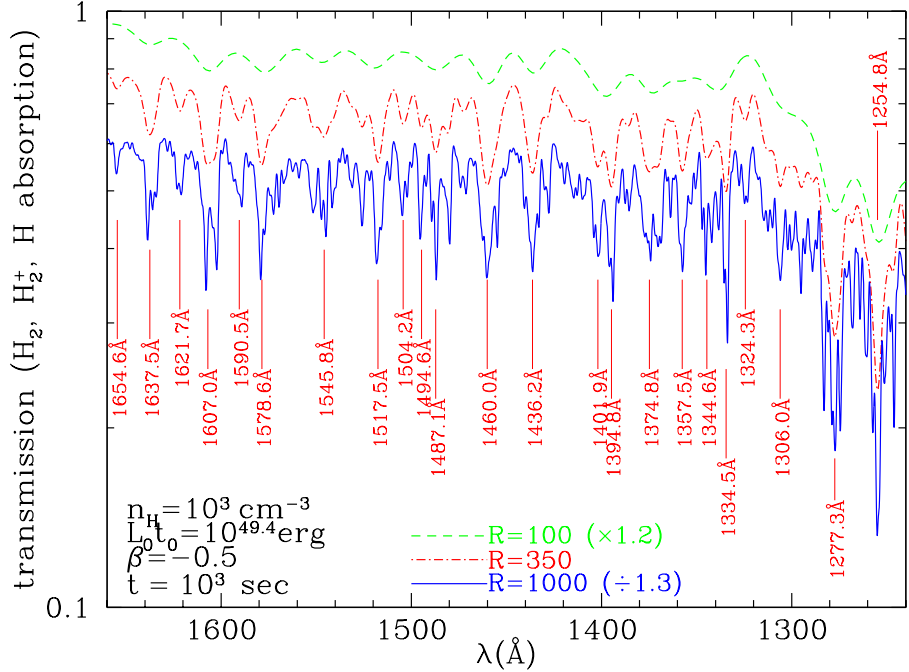}}
\figcaption{\label{fig:tran_n3_b5_t1000_R1003501000}
	{\footnotesize
	The $\lambda > 1110\Angstrom$ spectrum for the same conditions
	as Fig. \ref{fig:tran_n3_b5_t1000_R350} 
	at 
	spectral resolutions $R=100$, 350, and 1000.
	}
	}
\end{figure*}
\begin{figure*}[ht]	
	\centerline{\epsfig{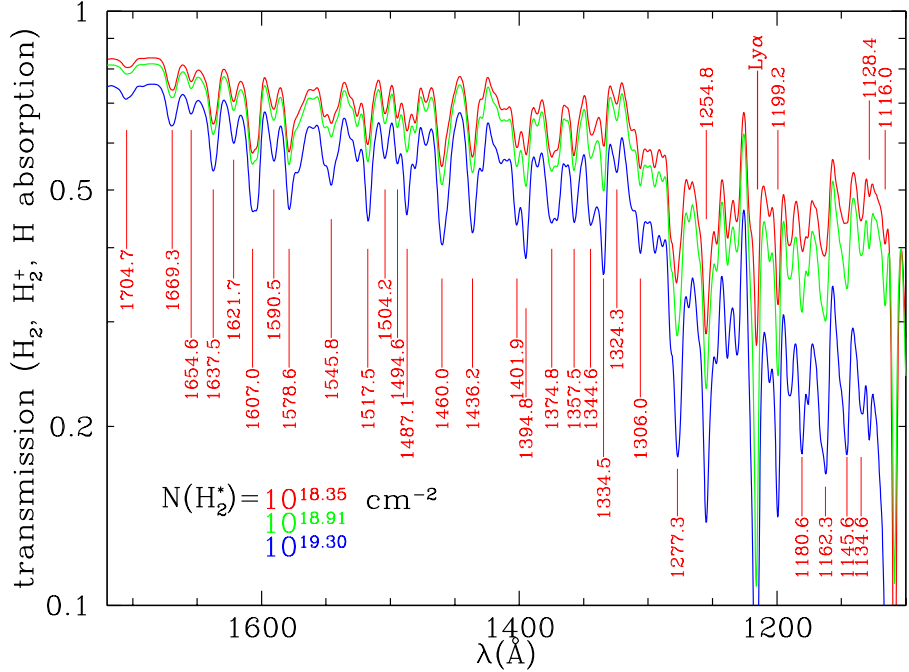}}
\figcaption{\label{fig:tran_for3Nstar}
	{\footnotesize
	The $\lambda > 1100\Angstrom$ transmission spectrum for
	$N(\HH^*)=10^{18.35}$, $10^{18.91}$, and $10^{19.30}\cm^{-2}$,
	at spectral resolution $R=350$.
	The absorption feature strengths are nearly unchanged due to
	saturation of the $\HH$ absorption lines.
	}
	}
\end{figure*}

\subsection{Vibrationally-Excited H$_2$
	\label{sec:N(H2star)}
	}

Following absorption of a Lyman or Werner band photon, about 85\% of
the time an $\HH$ molecule will undergo a spontaneous radiative decay
(in $\sim10^{-9}\s$) to a vibrationally-excited level of the
ground electronic state.
The vibrationally-excited $\HH$ will persist until it is either
photoionized or absorbs another Lyman or Werner band photon (on the
$\sim10^3\s$ timescales of interest here, spontaneous emission in the
quadrupole lines is negligible, as are collisional processes).
In Figure \ref{fig:N(H2star)}
we show the column density $N(\HH^*)$ of vibrationally-excited $\HH$ as
a function of time.
For all cases considered, $N(\HH^*)$ increases until the UV irradiation
ceases [unlike the $\HH^+$, which quickly stabilizes at a value
$N(\HH^+)\approx10^{16.7}\cm^{-2}$].
Photoionization acts to reduce 
the amount of vibrationally-excited $\HH$ present at any instant,
so GRB with softer UV-EUV spectra would be expected to have larger
column densities of vibrationally-excited $\HH$.
Our calculations confirm this: Fig.\ \ref{fig:N(H2star)}
shows that an optical transient with 
spectral index $\beta=-1$ (i.e., $L_\nu\propto\nu^{-1}$)
produces more vibrationally-excited $\HH$ than a burst with $\beta=-0.5$.

As we see below,
the large column densities of vibrationally-excited $\HH$ will produce
strong absorption lines.

\subsection{Emergent Spectrum
	\label{sec:emergent_spectrum}
	}
The radiation reaching us from the optical transient is filtered through
the dust and gas.
Unless and until the photoionization front reaches the edge of the gas
cloud, photoelectric absorption by H will impose a cutoff at
13.6~eV, with photoelectric absorption by $\HH$ further contributing
to absorption above 15.4~eV.
At X-ray energies $h\nu \gtsim 10~{\rm keV}$ the gas cloud may again
become transparent, allowing the X-ray flash and afterglow to be observed,
but our attention here will be limited to energies $h\nu < 13.6\eV$,
where there will be four main contributions to absorption:
dust, the Lyman and Werner bands of $\HH$,
photodissociation of $\HH^+$, and the Lyman lines of H.

The extinction by dust will affect whether the optical transient
is observable.
In principle there could be spectral features 
(e.g., the 2175$\Angstrom$ feature) 
in this absorption, but our understanding of the optical
properties of dust in these dense regions is such that no realistic
estimate is possible.
In Fig.\ \ref{fig:extincted_lightcurve} we show the emergent luminosity
after extinction by dust for ambient density $n_\rmH=10^3\cm^{-3}$.
Light curves are shown for various assumed cloud radii; for each case
the initial visual extinction $A_V$ to the GRB is indicated.
In the case of $R=1\times10^{19}\cm$ cloud radius, the ionization/dissociation
front reaches the cloud surface at $t\approx30\s$; after this time,
there will of course no longer be absorption features of $\HH$ appearing
in the spectrum of the GRB.
For $R> 1.4\times10^{19}\cm$, however, the ionization/dissociation front
is contained within the cloud, and the $\HH$ absorption lines will
persist in the spectrum of the GRB afterglow at late times.

For purposes of discussion we focus on our standard lightcurve with
$\beta=-0.5$ in a cloud of density $n_\rmH=10^3\cm^{-3}$.
By destroying the dust out to $\sim1.2\times10^{19}\cm$, the
GRB has reduced the extinction between us and the GRB by
$\Delta A\approx 5$~mag.  Thus if, for example, the dust in the cloud
had initially presented an extinction of, say, 8 mag to the GRB, 
the GRB afterglow would be subject to only 3 mag of foreground dust.

The $\HH^+$ column densities tend to be considerably smaller than
those of $\HH$ (see Figure \ref{fig:N(H2star)}), 
so we have not attempted to calculate the column densities
of the individual vibration-rotation levels and the
$X^2\Sigma_g^+\rightarrow C^2\Pi_u$ ultraviolet absorption lines out of
these levels.  Most of the electronic absorption by $\HH^+$ will 
be in $X^2\Sigma_g^+\rightarrow A^2\Sigma_u^+$ photodissociating
transitions, which contribute continuous absorption which we have
approximated by eq.\ (\ref{eq:sigma(H2+pd)}).

In Fig.\ \ref{fig:tran_n3_b5_t1000_R350} we show the transmission
of the foreground medium for the $n_\rmH=10^3\cm^{-3}$ cloud and
spectral index $\beta=-0.5$, for spectral resolution
$R=350$ characteristic of the grism on the Swift UVOT instrument.
Fig.\ \ref{fig:tran_n3_b5_t1000_R350} shows {\it only} the
absorption due to lines of $\HH$ and H, and the continuous absorption
of $\HH^+$.
Voigt profiles were used for both the H and $\HH$ lines.
Over the $\ltsim10^3\s$ timescales of interest here, the H and $\HH$ gas
is essentially collisionless.
For the $\HH$ lines 
we have therefore used a Doppler broadening parameter $b=3\kms$ characteristic
of ``microturbulence'' in the quiescent pre-GRB molecular gas.
The H atoms, on the other hand, will have a velocity distribution 
resulting from photodissociation of $\HH$ and $\HH^+$.
The mean kinetic energy per H atom is $\sim0.15\eV$ (Stephens \& Dalgarno 1973)
corresponding to a Maxwellian distribution with
$b\approx4\kms$; this will add in quadrature to the microturbulence, so we
take $b=5\kms$ for the H atoms.

The spectrum consists of literally thousands of 
saturated narrow lines, but
even at $R=350$ resolution,
a number of absorption features are conspicuous.
In Figure \ref{fig:tran_n3_b5_t1000_R350}
we have labelled 32 conspicuous features longward of $\lambda=1111\Angstrom$.
These features are all blends of a number of nearby lines.
For each feature in Table \ref{tab:features} 
we list the three strongest lines within
$\pm \lambda/(2R) = \pm1.3\times10^{-3}\lambda$ of the central wavelength
of each feature.
The strongest lines tend to be saturated
(e.g., the strongest line in the conspicuous 1277.3\AA\
feature is Lyman 8-6R(1) 1277.39\AA\, 
with a central optical depth $\tau_0=1150$ and
an equivalent width $W_\lambda/\lambda=8.3\times10^{-4}$).

Figure \ref{fig:tran_n3_b5_t1000_R1003501000} shows the transmission
spectrum observed with three different spectral resolutions:
$R=100$, 350, and 1000.
While the $R=1000$ spectrum of course shows the most structure,
we see that the $R=350$ resolution is quite well-matched to the widths of
the prominent absorption blends.
If the resolution is degraded to $R=100$, most of the absorption blends
still appear as well-defined minima, though a few do not (e.g.,
$\lambda=1621.7\Angstrom$.)
For $R=100$ observations, it would appear that a signal-to-noise ratio
of 10 or more would be sufficient to test for the presence or absence
of the stronger absorption features (e.g., 1607.0, 1460.0, 1401.9, 1277.3,
1254.8$\Angstrom$).

Because the lines are saturated, the overall spectrum is not sensitive
to the precise amount of vibrationally-excited $\HH$.
To see this, in Figure \ref{fig:tran_for3Nstar}
we show the $R=350$ spectrum for three different values of
the column density $N(\HH^*)$ of vibrationally-excited $\HH$,
ranging from $N(\HH^*)=10^{18.35}\cm^{-2}$ (the value at $t=2\s$ for
the $\beta=-0.5$, $n_\rmH=10^3\cm^{-3}$ case) to
$N(\HH^*)=10^{19.30}\cm^{-2}$ (the value at $t=10^3\s$ for
the $\beta=-0.5$, $n_\rmH=10^4\cm^{-3}$ case).
While the depth of the absorption features does increase with increasing
$N(\HH^*)$, the increase is
fairly modest despite the order-of-magnitude change in $N(\HH^*)$.

\begin{table}
\caption[]{$\lambda > 1110$\AA\ Absorption Features from UV-pumped
	H$_2$
	\label{tab:features}}
\begin{tabular}{llll}
\hline
$\lambda$(\AA)$^{a}$&\multicolumn{3}{c}{Three strongest contributing lines$^b$}\\
\cline{1-1}\cline{2-4}
1704.7&	3--13P(1) 1704.51&	3--13R(0) 1702.63&	3--13P(2) 1705.29\\

1669.3&	2--11R(1) 1667.46&	4--13R(1) 1667.28&	2--11P(1) 1670.40 \\

1654.6&	1--10R(1) 1654.99&	5--14P(1) 1654.19&	1--10R(0) 1655.40\\

1637.5&	4--12R(1) 1638.90&	3--11P(1) 1636.33&	4--12R(2) 1638.52\\

1621.7&	2--10R(1) 1620.74&	2--10P(1) 1623.51&	2--10R(0) 1621.05\\

1607.0&	6--13P(1) 1607.50&	5--12R(0) 1608.43&	5--12R(3) 1607.58\\

1590.5&	3--10R(1) 1588.76&	3--10P(1) 1591.31&	3--10R(0) 1588.98\\

1578.6&	7--13R(1) 1577.02&	0--8R(1) 1576.88&	7--13P(1) 1579.18\\

1545.8&	1--8R(1) 1544.90&	1--8P(1) 1547.54&	1--8R(0) 1544.94\\

1517.5&	0--7R(1) 1516.23&	2--8P(1) 1517.44&	0--7P(1) 1518.90\\

1504.2&	10--13P(1) 1504.15&	10--13P(2) 1504.92&	9--12P(1) 1503.78\\

1494.6& 1--7P(2) 1491.71&	1--7P(3) 1495.22&	11--14R(1) 1493.83\\

1487.1&	1--7R(1) 1486.63&	1--7P(1) 1489.08&	1--7R(0) 1486.53\\

1460.0&	0--6P(2) 1460.17&	4--8R(1) 1460.77&	0--6R(3) 1458.14\\

1436.2&	3--7P(1) 1435.05&	5--8R(1) 1436.09&	5--8P(1) 1438.02\\

1401.9&	2--6R(1) 1402.14&	0--5P(3) 1402.66&	2--6R(0) 1401.87\\

1394.8&	0--5R(1) 1393.97&	0--5P(1) 1396.23&	0--5R(0) 1393.73\\

1374.8&	10--9R(1) 1374.49&	1--5P(2) 1373.66&	10--9P(1) 1375.98\\

1357.5&	4--6R(1) 1355.56&	4--6R(2) 1357.35&	4--6R(2) 1356.85\\

1344.6&	2--5R(1) 1345.40&	2--5R(0) 1345.04&	7--7R(1) 1343.60\\

1334.5&	0--4R(1) 1333.80&	0--4R(0) 1333.48&	0--4P(1) 1335.87\\

1324.3& 3--5R(1) 1323.28&	3--5R(0) 1322.80&	3--5P(1) 1325.06\\

1306.0&	9--7R(1) 1306.21&	3--9R(1) 1305.59&	9--7P(1) 1307.58\\

1277.3&	0--3P(1) 1276.82&	0--3R(2) 1276.33&	0--3R(3) 1278.74\\

1254.8&	1--3R(1) 1253.94&	1--3R(0) 1253.50&	1--3P(1) 1255.68\\

1199.2&	1--2R(1) 1198.49&	1--2R(0) 1198.01&	1--2P(1) 1200.08\\

1180.6&	2--2R(1) 1180.42&	2--2P(1) 1181.89&	2--2R(2) 1181.93\\

1162.3&	1--4Q(1)$^c$ 1161.29&	0--1R(1) 1162.17&	3--2R(1) 1163.37\\

1145.6&	1--1R(1) 1144.71&	0--3Q(1)$^c$ 1145.90&	1--1P(1) 1146.16\\

1134.6&	2--1P(3) 1135.36&	2--4Q(2)$^c$ 1133.77&	15--5R(1) 1136.12\\

1128.4& 2--1R(1) 1128.22&	2--1R(0) 1127.68&	2--1R(2) 1129.74\\

1116.0&	1--3Q(1)$^c$ 1116.51&	1--3R(1)$^c$ 1114.93&	3--1R(3) 1116.66\\
\hline\\
\end{tabular}
\tablenotetext{a}{Wavelength 
of transmission minimum at resolution $R=350$.}
\tablenotetext{b}{All transitions are Lyman band unless otherwise indicated.}
\tablenotetext{c}{Werner band transition.}
\end{table}

\section{Observational Prospects: HETE and Swift
	\label{sec:obs_prospects}}

The presence of absence of the absorption features in Table \ref{tab:features}
will indicate whether or not GRBs occur in the vicinity of molecular gas.
The present paper has carried out simulations of the effect which 
the optical and
ultraviolet radiation from a GRB will have on nearby gas and dust, assuming
that the GRB progenitor is located in a uniform molecular medium of density
$n_\rmH=10^3$ or $10^4\cm^{-3}$.
It is important to realize, however, that vibrationally-excited $\HH$
can be produced even if the GRB is in a low-density region, but there
is molecular gas on the line-of-sight within a distance of a few
parsecs from the GRB.
Absorption features will be conspicuous for vibrationally-excited $\HH$ column
densities $N(\HH^*)> 10^{18}\cm^{-2}$.
This will be produced provided the total $\HH$ column 
exceeds $10^{21}\cm^{-2}$,
and the GRB flash includes a flux 
$(1/4\pi R^2)\int dt [\nu L_\nu/(h\nu)]_{1000\Angstrom} > F_{\min}$,
where $F_{\min}\approx 5\times 10^{18}\cm^{-2}$
of photons at $\lambda\approx1000\Angstrom$.
For the light curve (\ref{eq:L(t)}), this flux occurs at a distance
\beq
R < \left(\frac{L_0t_0}{2F_{\min}I_\rmH}\right)^{1/2}
\left(\frac{12.4}{13.6}\right)^{\beta/2}
\approx 110 \pc
\eeq
where we have taken $L_0t_0=2.5\times10^{49}\erg$ and $\beta=-0.5$.
Therefore even if the GRB occurs in a low-density region,
absorption by vibrationally-excited $\HH$ may be observable if
there is a molecular cloud on the line-of-sight within $\sim100$~pc
of the GRB.

Early detection and accurate positions for GRBs will enable spectroscopic
observations to be carried out while the fireball/afterglow is still bright.
The High Energy Transient Explorer (HETE) is expected to provide rapid
positions so that early ground-based observations may be possible for some
GRBs.
This may make allow ground-based telescopes to obtain $R > 100$
spectra of the afterglow 
for afterglows at $z > 1.25$ where the $\lambda=1600\Angstrom$ region
is redshifted into the U band.
For redshift $2.2 < z < 4.3$, the entire $1110-1705\Angstrom$ spectrum would
be observable from the ground with silicon CCD detectors.

The Swift Gamma Ray Burst Explorer mission, scheduled to be launched in 2003,
is expected to return positions within seconds for $\sim$1000 GRBs over its
3 year mission.
The UltraViolet and Optical Telescope on Swift will image
$\sim$300 GRBs within 70 sec of the burst, with 6500--1700$\Angstrom$
wavelength coverage.
For the brightest bursts, observations through a grism can be made,
yielding a $R\approx200$--400 spectrum over the
6500--1700$\Angstrom$ spectral range (Roming 2001).
The UV sensitivity means that rest wavelengths 1705 -- 1600$\Angstrom$ 
(a range which includes 6 of the absorption features in 
Table \ref{tab:features})
could be observed for GRBs at redshifts $z > 0.06$.  For GRBs at
$0.53 < z < 2.81$, the UVOT spectral range will include the entire
1705--1110$\Angstrom$ range where absorptions lines from vibrationally-excited
$\HH$ are conspicuous.

\section{Summary
	\label{sec:summary}
	}
The luminous optical transient associated with at least some GRBs will
have drastic effects on surrounding molecular gas and dust, if the GRB
is situated within a molecular cloud.
To illustrate the effects, we have adopted a model light curve for
the optical transient, and have considered power-law spectra
$L_\nu \propto \nu^{-0.5}$ and $L_\nu\propto \nu^{-1}$.
We have calculated the resulting photoioinization
and photodissociation of the surrounding gas, and thermal sublimation
of the dust, for two values of the gas density,
$n_\rmH=10^3$ and $10^4\cm^{-3}$.
Our principal results are as follows:

\begin{enumerate}
\item Dust will be destroyed out
to distances $R_d\approx 10^{19}\cm$ in clouds of density
$n_\rmH\ltsim 10^4\cm^{-3}$, confirming earlier estimates by WD00.

\item The ionization and dissociation fronts are merged
for spectra $L_\nu \propto \nu^{-0.5}$ and $\propto \nu^{-1}$.
The destruction of the $\HH$ is therefore due in part
to photodissociation, and in part to photoionization.

\item As proposed by Draine (2000), a substantial column density of
vibrationally-excited $\HH$ is created as long as the ionization-dissociation
front has not broken out of the molecular cloud.
This vibrationally-excited $\HH$ will produce strong absorption lines
in the $1705-1110\Angstrom$ region where normally
Lyman-$\alpha$ would be the only strong absorption line.

\item Spectra of GRB optical transients and afterglows should be compared
to the predicted absorption spectra shown in 
Figs.\ \ref{fig:tran_n3_b5_t1000_R350} -- \ref{fig:tran_for3Nstar}.
Even if the signal-to-noise ratio does not permit detection of
individual features, 
cross-correlation may make it possible to detect the
ensemble of lines between 1650 and 1110\AA.  Detection of such
absorption by vibrationally-excited $\HH$ would
be unequivocal evidence for association of the GRB with molecular gas.
The UVOT grism on the Swift Gamma Ray Burst Explorer mission may be
able to measure these features in the spectra of the brighter GRBs
at redshift $z\gtsim 0.1$.
\end{enumerate}
\acknowledgements
We thank
E. Roueff for making available $\HH$ data,
S. Hunsberger
for providing the specifications for the Swift UVOT instrument,
xxx for helpful discussions,
and
R. H. Lupton for availability of the SM software package.
This work was supported in part by
NSF grant AST-9988126.



\begin{thebibliography}{}
\bibitem[Abgrall et al.(1993a)]{ARL93a}
	Abgrall, H, Roueff, E., Launay, F., Roncin, J.-Y.,
	\& Subtil, J.-L.  1993a, A\&AS, 101, 273
\bibitem[Abgrall et al. 1993b]{ARL93b}
	Abgrall, H, Roueff, E., Launay, F., Roncin, J.-Y.,
	\& Subtil, J.-L. 1993b, A\&AS, 101, 323
\bibitem[Akerlof et al. (1999)]{ABB99}
	Akerlof, C.W., Balsano, R., Barthelmy, S. \etal\ 1999,
	Nature, 398, 400
\bibitem[Bates \& \"Opik (1968)]{BO68}
	Bates, D.R., \& \"Opik, U. 1968,
	J. Phys. B, 1, 543
\bibitem[Black \& Dalgarno 1976]{BD76}
	Black, J., \& Dalgarno, A. 1976, ApJ, 203, 132
\bibitem[Black \& van Dishoeck (1987)]{BvD87}
	Black, J., \& van Dishoeck, E. 1987, ApJ, 322, 412
\bibitem[Bloom et al.(1999)]{BKD99} 
	Bloom, J.S., Kulkarni, S.R., Djorgovski, S.G., \etal\ 1999, 
	Nature, 401, 453
\bibitem[B\"ottcher \etal.(1999)]{BDC99}
	B\"ottcher, M., Dermer, C.D., \& Liang, E.P. 1999, 
	A\&AS, 138, 343
\bibitem[Dalgarno  \etal. (1970)]{DHS70}
	Dalgarno, A., Herzberg, G., \& Stephens, T.L. 1970,
	ApJ, 162, L49
\bibitem[Draine, B.T. 2000]{Dr00}
	Draine, B.T. 2000, ApJ, 532, 273
\bibitem[Draine \& Bertoldi 1996]{DB96} 
	Draine, B.T., \& Bertoldi, F. 1996, ApJ, 468, 269
\bibitem[Fruchter (1999)]{Fr99} Fruchter, A. 1999, ApJL, 512, L1
\bibitem[Fruchter, Krolik, \& Rhoads (2001)]{FKR01}
	Fruchter, A., Krolik, J.H., \& Rhoads, J.E. 2001,
	ApJ, accepted [astro-ph/0106343]
\bibitem[Ghisellini \etal.(1999)]{GHC99}
	Ghisellini, G., Haardt, F., Campana, S., Lazzate, D.,
	\& Covino, S. 1999, ApJ, 517, 168
\bibitem[Guhathakurta \& Draine (1989)]{GD89}
	Guhathakurta, P., \& Draine, B.T. 1989,
	ApJ 345, 230
\bibitem[Habing (1968)]{Ha68}
	Habing, H.J. 1968, Bull. Astron. Inst. Netherlands, 19, 421
\bibitem[Israel \etal.(1999)]{IMC99}
	Israel, G.L., Marconi, G., Covino, S., \etal\ 1999,
	A\&A, 348, L51
\bibitem[Jansen \etal.(1995)]{JDB95}
	Jansen, D.J., van Dishoeck, E.F., Black, J.H., Spaans, M.,
	\& Sosin, C. 1995,
	A\&A, 302, 223
\bibitem[Lazzati etal 2001]{LCG2001}
	Lazzati, D., Covino, S., \& Ghisellini, G. 2001,
	MNRAS, submitted
	[astro-ph/0011443]
\bibitem[Lazzati etal 2001]{LPL2001}
	Lazzati, D., Perna, R., \& Loeb, A. 2001,
	MNRAS, submitted
	[astro-ph/0106202]
\bibitem[Lee etal (2001)]{LTV2001}
	Lee, B.C., et al.\ (2001), ApJ, accepted [astro-ph/0104201v2]
\bibitem[MacFadyen etal (1999)]{WF99}
	MacFadyen, A.I., Woosley, S.E., \& Heger, A. 2001, ApJ, 550, 410
\bibitem[Masetti etal (2001)]{MPP01}
	Masetti, N., et al.\ (2001), A\&A, accepted [astro-ph/0103296v2]
\bibitem[Osterbrock (1989)]{Os89}
	Osterbrock, D. E. 1989,
	Astrophysics of Gaseous Nebulae
	(Mill Valley: University Science Books)
\bibitem[Paczy\'nski (1998)]{Pac98}
	Paczy\'nski, B. 1998, ApJ, 494, L45
\bibitem[Paczy\'nski (1999)]{Pac99}
	Paczy\'nski, B. 1999, in The Largest Explosions Since the
	Big Bang: Supernovae and Gamma Ray Bursts,
	eds. M. Livio, K. Sahu, \& N. Panagia
	(Cambridge: Cambridge Univ. Press), in press
	[astro-ph/9909048v2]
\bibitem[Perna \& Loeb(1998)]{PL98}
	Perna, R., \& Loeb, A. 1998, ApJ, 501, 467
\bibitem[Ramirez-Ruiz, E. etal 2001]{RTB2001}
	Ramirez-Ruiz, E., Trentham, W., \& Blain, A.W. 2001,
	MNRAS, submitted
	[astro-ph/0103239]
\bibitem[Rodgers \& Williams (1974)]{RW74}
	Rodgers, C.D., \& Williams, A.P. 1974,
	J. Quant. Spectrosc. Radiat. Transfer, 14, 319
\bibitem[Roming (2001)]{Ro01}
	Roming, P. 2001, 
	Swift Specification for the Ultra Violet Optical Telescope,
	SWIFT-UVOT-002-R00,
	http://www.swift.psu.edu/uvot/doc/index.html
\bibitem[Roueff (1993)]{Ro93} Roueff, E., private communication
\bibitem[Smith \& Dwek (1998)]{SD98}
	Smith, R.K., \& Dwek, E. 1998, ApJ, 503, 831
\bibitem[Stanek \etal.(1999)]{SGK99}
	Stanek, K.Z, Garnavich, P.M., Kaluzny, J., Pych, W., \&
	Thompson, I. 1999, ApJ, 522, L39
\bibitem[Stanek \etal.(2001)]{SGJ01}
	Stanek, K.Z, et al.\ (2001),
	submitted to ApJ (Letters)
	[astro-ph/0104329v1]
\bibitem[Stephens \& Dalgarno (1973)]{SD73}
	Stephens, T.L., \& Dalgarno, A. 1973,
	ApJ, 186, 165
\bibitem[Sternberg (1988)]{St88}
	Sternberg, A. 1988, ApJ, 332, 400
\bibitem[Sternberg (1989)]{St89}
	Sternberg, A. 1989, ApJ, 347, 863
\bibitem[von Busch \& Dunn (1972)]{vBD72}
	von Busch, F., \& Dunn, G.H. 1972,
	Phys. Rev. A5, 1726
\bibitem[Vreeswijk \etal.(1999a)]{VGR99a}
	Vreeswijk, P.M., Galama, T.J., Rol, E., \etal, 1999a,
	GCN Circ. 310
\bibitem[Vreeswijk \etal.(1999b)]{VGR99b}
	Vreeswijk, P.M., Galama, T.J., Rol, E., \etal, 1999a,
	GCN Circ. 324
\bibitem[Waxman \& Draine (2000)]{WD00}
	Waxman, E.,\& Draine, B.T., 2000, ApJ, 537, 796
\bibitem[Williams et al (1999)]{WHP99}
	Williams, G.G., Hartmann, D.H., Park, H.S., et al. 1999,
	astro-ph/9912402
\bibitem[Yan, Sadeghpur, \& Dalgarno (1998)]{YSD98}
	Yan, M., Sadeghpour, H.R., \& Dalgarno, A. 1998, ApJ, 496, 1044
\end{thebibliography}
\end{document}